\documentclass{Rinton-P10x7}
\usepackage{amsfonts}

\begin{document}

\title{Quantum Computation with Harmonic Oscillators}

\author{Stephen D. Bartlett and Barry C. Sanders}

\address{Department of Physics, Macquarie University, Sydney, New
  South Wales 2109, Australia
  \\ E-mail: Stephen.Bartlett@mq.edu.au; Barry.Sanders@mq.edu.au
}

\author{Benjamin T.\ H.\ Varcoe}

\address{Max--Planck--Institut f\"ur Quantenoptik, 85748 Garching,
  Germany
  \\ E-mail: B.Varcoe@mpq.mpg.de}

\author{Hubert de Guise}

\address{Facult\'e Saint--Jean, University of Alberta, 8406 rue
        Marie--Anne Gaboury, Edmonton, Alberta, Canada T6C 4G9
  \\ E-mail: hdeguise@gpu.srv.ualberta.ca}


\maketitle

\abstracts{ By encoding a qudit in a harmonic oscillator and
  investigating the $d\to\infty$ limit, we give an entirely new
  realization of continuous--variable quantum computation.  The
  generalized Pauli group is generated by number and phase operators
  for harmonic oscillators.  We describe a physical realization in
  terms of modes in a microwave cavity, coupled via a standard Kerr
  nonlinearity.}

\section{Introduction}

The use of continuous--variable~(CV) quantum computing allows
information to be encoded and processed much more compactly and
efficiently than with discrete--variable~(qubit) computing.  With CV
realizations, one can perform quantum information processes using
fewer coupled quantum systems: a considerable advantage for the
experimental realization of quantum computing.  The rapidly developing
field of CV quantum information theory has applications to quantum
error correction~\cite{Got00}, quantum cryptography~\cite{Ral00} and
quantum teleportation~\cite{Bra98}, including an experimental
realization of CV quantum teleportation~\cite{Fur98}.

The most common proposed realization of CV quantum computation at
present employs position eigenstates as a computational
basis~\cite{Got00}; these states are approximated experimentally using
highly squeezed states~\cite{Bra98}.  Other CV realizations are
described by a generalized Pauli group generated by the number
operator $\hat{N}$ and a phase operator $\hat{\theta}$, and
computational bases given by either harmonic oscillator number
eigenstates or phase eigenstates~\cite{Bar00}.  These realizations are
obtained formally by taking the $d \to \infty$ limit of the qudit, the
$d$--dimensional generalization of the qubit, and are important for
five key reasons: (\emph{i})~these CV realizations are entirely
distinct from the position eigenstate computational basis realization,
both in terms of the computational basis and in terms of the SUM gate;
(\emph{ii})~the SUM gate employs a standard Kerr optical nonlinearity
to couple two modes; (\emph{iii})~these realizations give natural
extensions of the qubit--based (discrete--variable) Pauli group, with
a well--defined limiting procedure; (\emph{iv})~this CV quantum
computation presents an appealing realization in terms of coupled
modes in a microwave cavity using the powerful methods of state
preparation~\cite{Vog93} in such cavities; and (\emph{v})~these
realizations give a new implementation of the well--studied phase
operator~\cite{Peg97}.

\section{CV Quantum Computation using Harmonic Oscillators}

To describe CV quantum computation, one requires an
infinite--dimensional Hilbert space $\mathcal{H}$ and a representation
of the generalized Pauli group~\cite{Got00} that serves as a group of
unitary transformations on $\mathcal{H}$.  By defining $\mathcal{H}$
to be the Hilbert space for a harmonic oscillator, one can define
specific representations of the generalized Pauli group in terms of
operators that act on the harmonic oscillator Hilbert space.

The natural generalization of the Pauli group~\cite{Bar00} is the
group generated by the number operator $\hat{N}$ and the
(Pegg--Barnett~\cite{Peg97}) phase operator $\hat{\theta}$.  Note that
it is convenient to work with the finite--dimensional Hilbert space
$\mathbb{H}_d$ of boson number not greater than $d-1$.  On this
Hilbert space, the phase operator is well--defined, and the system
incorporates a finite energy cutoff and corresponding phase
resolution.

With this generalization of the Pauli group, there are two natural
computational bases.  The first, the \emph{number state basis},
consists of the harmonic oscillator energy eigenstates
\begin{equation}
  \label{eq:NumberStateBasis}
  \hat{N} |n\rangle = n |n\rangle \, , \quad n=0,1,\ldots, d-1 \, .
\end{equation}
The second basis, the \emph{phase state basis}, consists of
eigenstates of the phase operator
\begin{equation}
  \label{eq:PhaseStateBasis}
  \hat{\theta} |\phi\rangle = \phi |\phi\rangle \, , \quad
  \phi=0,\frac{2\pi}{d}, \ldots,
  \frac{2(d-1)\pi}{d} \, .
\end{equation}
These bases are ``dual'' in the sense that $\langle \phi | n \rangle =
\exp(i\phi n)$.

To perform universal CV quantum computation~\cite{Llo99}, it is
necessary to be able to: (\emph{i})~initially prepare a qudit in an
arbitrary state for computation; (\emph{ii})~realize an arbitrary
unitary transformation on a single qudit;  (\emph{iii})~have a controlled
two--qudit interaction gate such as the SUM gate~\cite{Got00}
\begin{equation}
  \label{eq:SUMGate}
  {\rm SUM}: \ |s_1\rangle_1\otimes|s_2\rangle_2 \mapsto
  |s_1\rangle_1\otimes|s_1+s_2\rangle_2 \, ;
\end{equation}
and (\emph{iv}) perform von Neumann measurements in the computational
basis.

\section{Proposed Realization as Coupled Modes of a Microwave Cavity}

One approach to physically realizing CV (qudit--based) quantum
computation is by using a multimode microwave cavity.  A single qudit
is realized as the state of a (longitudinal) mode, and qudit--qudit
interactions are performed by appropriately coupling different modes.
By using a microwave cavity, one can take advantage of the recent
technical advances of the micromaser: a micromaser provides a physical
realization of the Jaynes--Cummings model, which describes the coupled
system consisting of a single two--level atom and a single mode of the
radiation field.  Single atom interactions enable state preparation,
qudit transformations, two--qudit coupling, and state measurement.
Current high--$Q$ microwave cavities allow photon lifetimes up to
0.3~s~\cite{Var00}, several orders of magnitude larger than the atom
interaction time.

State preparation in a microwave cavity is highly developed, both
theoretically and experimentally.  Proposals~\cite{Vog93} for
preparation of an arbitrary state of the radiation field involve
injecting a sequence of atoms and are conditional on measurement of
the output atoms; a deterministic approach to the preparation of Fock
states (not involving conditional measurements) has been
experimentally realised~\cite{Var00}.  Recently, it has been
proposed~\cite{Wel00} that any state of the radiation field can be
prepared with arbitrarily high fidelity using a sequence of entangled
atoms; this technique has the important features that the resulting
state of the field is not entangled with that of the atoms, and that
the method is independent of the initial state of the field.

An arbitrary unitary transformation on a single qudit (single mode),
to any desired precision, can be performed \emph{efficiently} using a
combination of phase--space displacements, squeezing, and a nonlinear
Kerr effect~\cite{Llo99}.  By an appropriate combination of these
transformations, one can approximate (to arbitrary accuracy) any
polynomial Hamiltonian in $\hat{a}^\dagger$ and $\hat{a}$.

Linear transformations of a cavity mode state may be performed by
coupling the mode to an external monochromatic coherent field via
transmission through one of the mirrors~\cite{Bod98}.  A strong
coherent field is directed into the cavity through a
low--transmissivity mirror, and the result on the mode state is that it
is displaced; i.e., the field state is acted upon by a unitary Glauber
displacement operator.  The choice of frequency for the incoming
coherent field determines which of the longitudinal field states (each
with a different frequency) is displaced.

Squeezing the field state requires a nonlinear process, as does the
nonlinear Kerr transformation of the field state.  The squeezing
transformation may be achieved via a~$\chi^{(2)}$ nonlinearity and the
latter by a~$\chi^{(3)}$ nonlinearity.  A nonlinearity is effected by
passing atoms through the cavity one at a time: during the time of
passage, the field states undergoes a nonlinear evolution.  An
effective nonlinearity is obtained by adiabatically eliminating the
slowly varying atomic degrees of freedom.  The field in the cavity may
be squeezed by injecting two--level atoms into a micromaser
cavity~\cite{Kim91} or three--level atoms~\cite{Ors92}.  By choosing
atomic levels and motional parameters carefully, the squeezing
(Bogoliubov) transformation can be approximated.  The Kerr, or
$\chi^{(3)}$, nonlinearity is generally smaller, but it is possible to
employ a four--level atom, passing through the cavity, to obtain a
large Kerr nonlinearity~\cite{Reb99}.  Thus, by directing atoms
through the cavity, unitary $\chi^{(2)}$ and $\chi^{(3)}$ evolutions
particular field states can be effected.  The Kerr nonlinearity can
also effect a cross--phase modulation to perform the micromaser
equivalent of the optical cavity quantum electrodynamics conditional
phase shift~\cite{Tur95}.

For quantum computation, we must also realize a gate that performs a
two--qudit interaction.  Consider two oscillators coupled by the
four--wave mixing interaction Hamiltonian $\chi \hat{N}_1 \hat{N}_2 =
\chi \hat{a}_1^\dagger \hat{a}_1 \hat{a}_2^\dagger \hat{a}_2$.  This
Hamiltonian for an optical system describes a four--wave mixing
process in which $\chi$ is proportional to the third--order nonlinear
susceptibility~\cite{Mil83}.  Let oscillator $1$ be in a state $|s_1
\rangle_1$ encoded in the number state basis, and let oscillator $2$
be in a state $|s_2 \rangle_2$ encoded in the phase state basis.  This
interaction Hamiltonian generates the transformation
\begin{equation}
  \label{eq:SUMHamiltonian}
  e^{-{\rm i} \chi \hat{N}_1 \hat{N}_2 t} |s_1 \rangle_1 \otimes |s_2
  \rangle_2 = |s_1 \rangle_1 \otimes |\chi t s_1 + s_2 \rangle_2 \, .
\end{equation}
Thus, with time $t=\chi^{-1}$, this Hamiltonian generates the SUM
transformation.

Quantum computation with multiple qudits could be performed by
coupling several modes in a single cavity; each mode realizes a single
qudit.  Modes are coupled via a SUM interaction of the time described
above.  Note that the control qudit for the sum operation must be
encoded in the number state basis, and the target qudit must be in the
phase state basis

Measurements in the number state computational basis can be performed
using a QND measurement of photon number~\cite{Hol91}.  Atoms
passing through the cavity have their momentum coupled to the
radiation field; the momentum distribution of outgoing atoms yields a
measurement of photon number.

In summary, we have presented a new form of continuous variable
computation in terms of number and phase operators, and describe a
realization in terms of coupled modes of a microwave cavity using
linear transformations, squeezing, and nonlinear Kerr media.  This new
approach has the advantage over position--eigenstate CV computation in
that the computational basis states, for large but finite $d$, are
well--defined and obtainable, and do not require
``infinite--squeezing'' of Gaussian wavepackets.

\section*{Acknowledgments}

This project has been supported by an Australian Research Council
Large Grant and by a Macquarie University Research Grant.  We
acknowledge helpful discussions with Samuel Braunstein and Terry
Rudolph.

\end{document}